\documentclass[manuscript,nonacm]{acmart}

\AtBeginDocument{%
  \providecommand\BibTeX{{%
    \normalfont B\kern-0.5em{\scshape i\kern-0.25em b}\kern-0.8em\TeX}}}

\setcopyright{none}
\copyrightyear{2023}
\acmYear{2023}
\acmDOI{XXXXXXX.XXXXXXX}

\acmConference[CSCW'23 Workshop]{A Toolbox of Feminist Wonder Workshop}{October 15, 2023}{Minneapolis, MN}
\acmPrice{15.00}
\acmISBN{978-1-4503-XXXX-X/18/06}

\begin{document}

\title{Frameworking for a Community-led Feminist Ethics}

\author{Ana O. Henriques}
\email{ana.gfo.henriques@campus.ul.pt}
\author{Hugo Nicolau}
\email{hugo.nicolau@tecnico.ulisboa.pt}
\affiliation{%
  \institution{Interactive Technologies Institute / LARSyS}
  \city{Lisbon}
  \country{Portugal}
}
\author{Kyle Montague}
\email{kyle.montague@northumbria.ac.uk}
\affiliation{%
  \institution{Northumbria University}
  \city{Newcastle}
  \country{United Kingdom}
}


\begin{abstract}

This paper introduces a relational perspective on ethics within the context of Feminist Digital Civics and community-led design. Ethics work in HCI has primarily focused on prescriptive machine ethics and bioethics principles rather than people. In response, we advocate for a community-led, processual approach to ethics, acknowledging power dynamics and local contexts. We thus propose a multidimensional adaptive model for ethics in HCI design, integrating an intersectional feminist ethical lens. This framework embraces feminist epistemologies, methods, and methodologies, fostering a reflexive practice. By weaving together situated knowledges, standpoint theory, intersectionality, participatory methods, and care ethics, our approach offers a holistic foundation for ethics in HCI, aiming to advance community-led practices and enrich the discourse surrounding ethics within this field.

\end{abstract}

\ccsdesc[500]{Human-centered computing~Human computer interaction (HCI)}
\ccsdesc[500]{Human-centered computing~HCI theory, concepts and models}

\keywords{feminism, ethics, feminist ethics, design justice, community-led design}


\maketitle


\section{Introduction}

Feminism, especially within academia, is typically held as a domain of critical theory aimed at analyzing the systemic and manifold ways gendered oppression manifests. It is plural in both construction and presentation but has as its key concerns "issues such as agency, fulfillment, identity, equity, empowerment, and social justice" \cite{Bardzell2010} (p. 1302). On the other hand, the field of digital civics aims to empower citizens and non-state actors to co-create, take an active role in shaping agendas and move from transactional to relational service models due to the potential of such models to reconfigure power relations between citizens, communities, and institutions \cite{Vlachokyriakos2016}. Further, within the purview of digital civics, community-led design is a movement focused on reframing the approach to co-design with a specific focus on empowering communities to catalyze their own needs/context-based solutions \cite{Sanders2008}. Given the overlap in intention, we believe these missions to be intimately entwined with those of feminism. Indeed, this paper will outline our work on developing an ethics for digital civics that explicitly and purposely draws from feminist theory and praxis.

Within HCI, the discussion surrounding ethics typically focuses on machine ethics and bioethics principles rather than considerations about and between people. Additionally, all ethics work in HCI is addressed to other researchers, even that which draws from a community-led approach to design and research. Such work also implies by omission that the default expectation is that researchers do not already belong to these communities, which begs considerations on power and justice within the academic space.

 We thus propose a more relational and reflexive approach to ethics. We do so by proposing a process of frameworking — rather than a fixed framework — for feminist ethics with the primary goal of developing a processual ethics toolkit that communities can use independently.


In so doing, we see an inextricable link between the feminist ethos and that of our work, which will be described below, along with the most relevant feminist epistemologies, methods, and methodologies from which we draw. This structure was modeled after Bardzell and Bardzell's paper \cite{Bardzell2011} outlining a feminist methodology for HCI — symbolizing both that we hope to add our voices to that call and that we recognize the importance of the seminal work done already. 

Ethics, especially within the context of HCI tends to be prescriptive, typically following the form of set guidelines or codes of conduct, rather than processual, where it might take on forms more akin to Komesaroff's micro-ethics approach \cite{Komesaroff1995} but without its strict reliance on static codes \cite{Bittner2005}. This static nature of hitherto ethics applications leaves little room for contingency, variation, and social dynamics — indeed, for life. This is where we would like to intervene. We propose a more dynamic form of ethics, community-led and continuous, which accounts for local contexts and is inherently cognizant of how power dynamics can influence decisions; and, in turn, society \cite{Lynch2016}.


\section{Frameworking for a Community-led Feminist Ethics}


Ethical deliberations are concerned with everything from the individual to the collective. They are inherently shaped by moral agendas, which are necessarily contingent. This is important to note because if gone unexamined, these agendas can produce harmful outcomes that further oppressive value systems \cite{Henriques2023}. This is especially true when considering unbalanced power/knowledge dynamics \cite{Foucault1980} regarding marginalized epistemologies. Ethical processes of deliberation must take into consideration the complexities surrounding exercises of power. This is precisely why an intersectional feminist framework is particularly apt, given its inherent concern for power differentials, which "helps in contextualizing the pervasive silencing, absence, difference, and gendered oppression in all its intersecting forms while remaining conscious of the relationality inherent to any deliberation process" \cite{Henriques2023} (p. 2).

\subsection{Aim}

Ruth Levitas presents a framework for societal reconstitution through three distinct yet interrelated modes. The first, archaeological, involves linking concepts and images within political agendas and policies. The second, ontological, examines and questions the prevailing values and knowledge systems that shape a given society. The last, architectural, focuses on empowering individuals to envision and conceptualize alternative possibilities for the future \cite{Levitas2013}. Though Levitas's work was on utopianism, we find within it many parallels to the aims of this work. The utopian tradition is predicated on the concept of hope for the essential ideation of betterment, which constitutes the basis of any utopian project \cite{Anderson2006}. And hope, argues José Muñoz, is also praxis — a critical methodology which informs "a backward glance that enacts a future vision" \cite{Munoz2009} (p. 16), even when it is difficult. Indeed, especially when it is.

Drawing from these foundations, we build on our prior work \cite{Henriques2023} to develop a robust and multidimensional adaptive model for community-led ethics in HCI design that acknowledges the necessity of a layered analysis and incorporates an intersectional feminist ethical framework as an integral component thereof. To do so, we are working on developing a framework primarily focused on independent use by communities in addition to a flexible toolkit for the explicit incorporation of ethical considerations in community-led design projects \cite{Chock2020} on a case-by-case basis. We purposely focus on community-led design as a movement that intends to shift from \textit{designing for/with} to a \textit{designing by} approach. Key to this design idea is the concept of empowering communities with the skills necessary to innovate and create solutions for themselves. Community-led design is an approach in which the co-design process \cite{Sanders2008}, not just the outcome, is developed through collaborations with community members who will be directly impacted by the design.

\subsection{Epistemologies}

An epistemology is a theory of knowledge. It is, in essence, how we know, and for whom as well as how we produce knowledge \cite{Bardzell2011}. Standpoint theory, along with Donna Haraway's concept of situated knowledges, is a central element of feminist epistemology. Situated knowledges emphasize the idea that knowledge is partial and contextually situated rather than universal and transcendent \cite{Haraway1988}. Historically, assumed objectivity has been portrayed as an impartial standpoint, seemingly neutral, but it often perpetuates power dynamics rooted in the dominance of masculinity, whiteness, and normative gender norms, among other axes in the Matrix of Domination \cite{Collins1990} \cite{Chock2020}. In contrast to the dominant epistemological tradition, feminist epistemologies focus on the influence of social positioning. Standpoint theory addresses this by emphasizing that a standpoint is not inherited but rather achieved through active political engagement with feminist agendas \cite{Harding2004}. Feminist standpoints thus allow us to step outside the boundaries of conventional approaches and encourage us to seek alternatives. By embracing a solid feminist research ethic, we can also develop greater confidence in the tools we employ in our projects, which ultimately elevates the quality of our research \cite{Henriques2023}.

Though distinct from epistemologies per se, we feel that including some theoretical frameworks\footnote{Very simply put, epistemology is concerned with how knowledge is produced. A theoretical framework is what one comes to understand \cite{Tennis2008}.} is still appropriate within this section. Drawing on a framework of intersectionality \cite{Crenshaw1989} \cite{Collins1990}, which includes disability justice, and queer and post-colonial analyses of power, we are also including principles of justice within our work on ethics. We are particularly interested in those of Nancy Fraser \cite{Fraser2009} and Amartya Sen \cite{Sen2012}, whose work on justice requirements and inclusion is invaluable in shaping our ethics toward equity in HCI design — which can be defined as "the acknowledgement of oppression stemming from social systems" and "involves designing with the intent to address such oppression by valuing community perspectives and redistributing design power to marginalized communities" \cite{Petterson2023} (p. 1).

\subsection{Methods}

Methods are techniques for data collection and analysis \cite{Bardzell2011}. Methods often employed by feminist scholars and practitioners are largely qualitative and include in-depth interviews, discourse analysis, observation, reflection workshops, focus groups \cite{Wilkinson1998}, or group-level assessments \cite{Vaughn2014}, all of which we are utilizing to inform the design of the ethical framework. This emphasis on gathering information on quotidian experiences is a hallmark of the feminist ethos, and crucial to our work (e.g., \cite{Neto2021}). Perhaps for that reason, feminist researchers frequently draw from participatory methods to empower people to mold the research \cite{Cook1986} cited in \cite{Bardzell2011}. Specifically, we intend to iterate on the framework itself through participatory methods to develop a toolkit for later community-led use.

We are co-designing the toolkit through methods rooted in a value-sensitive design approach that draws from the previously developed framework and is informed by Petterson et al.'s review of toolkits geared at promoting equity \cite{Petterson2023}. However, we intend to leave space for communities to determine and prioritize their own needs and values. As such, we plan on iterating on the \textit{schnittmuster} method for designing adaptable toolkits as an approach specifically developed to account for contextual variation \cite{Meissner2018}. To do so, we believe a method such as (Counter)storytelling to be a constructive glue to hold our \textit{schnittmuster} together. Indeed, (re)telling stories has been a crucial element in feminist work \cite{Lauretis1986}, and it has even been shown to promote increased involvement and interest in the research process \cite{Lennie1999}. Counterstorytelling in particular, which finds its roots in Critical Race Theory, aims to foster personal growth and challenge dominant paradigms, and has been found to foster a greater understanding of intersectional marginalized identities \cite{Wagaman2018}.

\subsection{Methodologies}

Methodology is the link between epistemology and method. It is the practical implementation of an epistemology through the curation of appropriate methods, which should adequately and explicitly reflect the theoretical considerations informing any given research context \cite{Bardzell2011}. Specifically, feminist methodologies strive to represent the diversity of human experiences and redirect knowledge production away from control and toward nurturing \cite{Sprague2005}. This, as argued in \cite{Bardzell2011}, "is especially urgent" (p. 681).

Indeed, we consider an approach rooted in care ethics as the basis of our ethics framework. Largely modeled after Joan Tronto's work exploring the intersections between care ethics, feminist theory and politics, we too understand ethics as an \textit{exercise} in care, defined broadly as "a species of activity that includes everything we do to maintain, contain, and repair our 'world' so that we can live in it as well as possible. That world includes our bodies, ourselves, and our environment" \cite{Tronto1990}. Affect, as Brian Massumi argues, holds the potential to disrupt and challenge conventional power/knowledge structures, making it a powerful force for political mobilization and resistance against oppressive systems \cite{Massumi2015}. As such, we contend that a reframing of ethics as that which occurs between people could facilitate a more reflexive practice — situational and hence political.

Moreover, we look upon Bardzell's description of a generative approach to the integration of feminism into HCI, in which she describes six potential qualities which ought to underpin that process — pluralism, participation, advocacy, ecology, embodiment, and self-disclosure \cite{Bardzell2010}. Beyond our attempts to integrate them into our work, we see these also as a self-assessment methodology. This is how we understand Costanza-Chock's work on Design Justice. More than a theoretical exploration of how design can help empower marginalized communities, dismantle structural inequality, and foster collective liberation \cite{Chock2020}, we see it also as a set of evaluative case studies for how community-led design practices can elevate political agendas of social reform — something which we perceive to be inextricable from a feminist ethics.

\section{Conclusion}

Toward our goal of developing a community-led process of frameworking for ethics, we believe that feminist theory and praxis must be woven into its very fabric. Indeed, we explore how feminist epistemologies and methods can inform our methodology, through which we hope to challenge conventional notions of objectivity by foregrounding social positioning and power dynamics. Situated knowledges and standpoint theory advocate for a contextual and situational understanding of knowledge, shaped by social and political engagement. The discussion also extends to intersectionality, incorporating disability justice, queer, and post-colonial analyses to enrich ethical considerations. Feminist-informed methods foster empowerment through participatory engagement geared toward justice and equity. The intricate application of a value-sensitive design approach, the \textit{schnittmuster} method, and the integration of (counter)storytelling illuminate our methodological landscape. Feminist methodologies' dedication to embracing diverse human experiences while nurturing knowledge creation is precisely what we aim for, and embracing a care ethics framework, as well as qualities like pluralism and participation, underscores our holistic approach to feminist HCI.

It should not be left unsaid, however, that the outline we propose here is only one way of approaching the problem. Feminism is inherently plural, hence its value in works of dissent. In keeping with Bardzell and Bardzell's call for a feminist HCI methodology, we wish to build something adaptable rather than dogmatic — alive like we are. Because, ultimately, any work we do should strive to improve our collective life experiences in a social world. And feminist ethics will, inherently, observe the power differentials which permeate any design and research process, and is thus more likely to conduce actual transformative potential.

As such, we hope our contributions will have value for anyone who interacts with our work. Even though we are primarily addressing local communities, we believe this interests the broader academic community, especially those studying ethics in HCI and those working to further community-led practices and/or embedded research. We estimate to be able to contribute with the framework itself, findings from our participatory methods, and our \textit{schnittmuster} toolkit, in addition to its use. This is still a work in progress; thus, we would gladly welcome any feedback and suggestions.

\begin{acks}
This work was supported by the European project DCitizens (GA 101079116) and the FCT project UIDB/50009/2020. We would like to acknowledge the invaluable insights that working alongside the Balcão do Bairro project and the Portuguese Council for Refugees have given us. We thank all the authors we reference and all those on whom we stand.
\end{acks}

\bibliographystyle{ACM-Reference-Format}
\bibliography{references}

\end{document}